\documentclass[11pt,twocolumn]{article}

\usepackage{amsfonts,amsmath,amssymb,bm,abstract,balance}

\title{\textbf{Pairing of charged particles in a quantum plasmoid}}

\author{Maxim Dvornikov
\\
\small{Institute of Physics, University of S\~{a}o Paulo,} \\
\small{CP 66318, CEP 05315-970 S\~{a}o Paulo, SP, Brazil and} \\
\small{Pushkov Institute of Terrestrial Magnetism, Ionosphere} \\
\small{and Radiowave Propagation (IZMIRAN),} \\
\small{142190 Troitsk, Moscow Region, Russia} \\
\small{E-mail: maxim.dvornikov@usp.br}}

\date{}

\begin{document}

\twocolumn[\maketitle
\begin{onecolabstract}
We study a quantum  spherically symmetric object which is based on
radial plasma oscillations. Such a plasmoid is supposed to exist
in a dense plasma containing electrons, ions, and neutral particles.
The method of creation and annihilation operators is applied to quantize
the motion of charged particles in a self-consistent potential. We
also study the effective interaction between oscillating particles
owing to the exchange of a virtual acoustic wave, which is excited
in the neutral component of plasma. It is shown that this interaction
can be attractive and result in the formation of ion pairs.
We discuss possible applications of this phenomenon in astrophysical
and terrestrial plasmas.
\end{onecolabstract}]


%
%
%
%
%
%
%
%


\section{Introduction}

The studies of quantum effects in the dynamics of charged particles
in dense plasmas is a rapidly developing branch of the modern plasma
physics~\cite{Man05}. Numerous effects such as the behavior of
quantum dots~\cite{BalBonLeeStaDah09}, the exciton dynamics~\cite{Ark05}
etc, unusual for classical systems, were recently reported to exist
when quantum dynamics is taken into account. Note that quantum effects
in astrophysical plasmas are also important in the evolution of compact
stars (see, e.g., Ref.~\cite{Mel08}). However, as it was mentioned
in Ref.~\cite{VlaTys11}, still there is a lack of understanding
how to correctly account for the quantum dynamics in plasma physics.
In particular, a careful analysis of applicability of any of the approaches
for the description of quantum plasmas should be made.

In the present work we shall discuss the quantum dynamics of a certain
class of plasma objects. We shall study a spherically symmetric plasmoid
based on radial oscillations of charged particles. Note that such
a configuration of plasma oscillations was previously analyzed in
connection to the studies of the Langmuir waves collapse~\cite{Zak72}.
We also mention that spherically symmetric plasma oscillations can
be a theoretical model of stable natural plasmoids~\cite{Zhe90,Fed99,DvoDvo06,Shm03,Ten11}.

We start in Sec.~\ref{sec:MODQUANTPLAS} with the formulation of
a model for a quantum spherical plasmoid based on radial plasma oscillations.
Then, in Sec.~\ref{sec:QUANTEL}, the motion of charged particles
is quantized using the general ideas of Sec.~\ref{sec:MODQUANTPLAS}.
In Sec.~\ref{sec:QUANTPH} we carry out a quantization of acoustic
waves which are excited in the neutral component of plasma existing
in our system. We describe the effective interaction between charged
particles owing to the exchange of an acoustic wave and analyze when
this interaction can be attractive in Sec.~\ref{sec:INTELACOU}.
In Sec.~\ref{sec:DIAGHAM} we study the possibility of the formation of
bound states of oscillating charged particles in a spherical plasmoid.
Finally, in Sec.~\ref{sec:APPL} some applications are considered.
In Sec.~\ref{sec:CONCL} we briefly summarize our results.

\section{Model of a quantum plasmoid\label{sec:MODQUANTPLAS}}

One of the main approaches for the quantitative studies of quantum effects in dense
plasmas consists in accounting for the quantum pressure term in the
Euler equation, i.e., for instance, the pressure becomes~\cite{key-01},
\begin{equation}
p\to p-\frac{\hbar^{2}}{2m^{2}\sqrt{n}}\nabla^{2}\sqrt{n},\label{eq:Bohmpress}
\end{equation}
where $m$ is the particle mass and $n$ is the number density. Note
that the modification~(\ref{eq:Bohmpress}) is likely to account
for small corrections to the dynamics of classical plasmas. For example,
there is an additional term in the dispersion relation for Langmuir
plasma waves, $\omega^{2}=\omega_{\mathrm{cl}}^{2}(\mathbf{q})+\hbar^{2}\mathbf{q}^{4}/4m_{e}^{2}$
(see, e.g., Ref.~\cite{BomPin53}), where $\omega$ is the frequency
of plasma oscillations, $\omega_{\mathrm{cl}}(\mathbf{q})$ is the
classical dispersion relation, $m_{e}$ is the electron mass, and
$\mathbf{q}$ is the wave vector. Thus this approach for the description
of the plasma dynamics seems to be the quasi-classical method for
the studies of quantum effects in plasma physics. Therefore, to describe
the dynamics of a strongly correlated system of charged particles,
this method is unlikely to be valid and the new approach has to be
developed.

If we are interested in the studies of a quantum object based on oscillations
of charged particles in plasma~\cite{DvoDvo06}, the main problem in
this description is the construction of a ground state of this many
body system. The ground state based on the particles wave functions
taken in the form of plane waves, as it is typically made in the condensed
matter physics, is inappropriate since it does not fully reflect the
dynamical features of the system. Thus we should build the ground
state using the wave functions corresponding to the oscillatory motion
of charged particles.

The dynamics of plasma oscillations on the classical level was studied
in Ref.~\cite{Daw59} using the Lagrange variables method. In this
formalism, the time evolution of a particular charged particle is
analyzed. It was found that the separation of positive and negative
charges in plasma results in the appearance of oscillations of lightest
charged particles, i.e. electrons, with the Langmuir frequency $\omega_{e}=\sqrt{4\pi e^{2}n_{e}^{(0)}/m_{e}}$,
where $e>0$ is the proton charge and $n_{e}^{(0)}$ is the equilibrium
number density of electrons. It means that the equation of motion
of an electron has the form,
\begin{equation}\label{eq:classeqmot}
  \ddot{\mathbf{r}} = -\omega_{e}^{2}\mathbf{r}.
\end{equation}
Thus one can suggest that the perturbed motion of a single electron
in plasma is governed by the quadratic potential $V(\mathbf{r})=m_{e}\omega_{e}^{2}r^{2}/2$.
In this case one can reproduce the classical equation of motion~(\ref{eq:classeqmot}).

We should also mention that the idealized picture described above
when only electrons participate in plasma oscillations is never implemented.
In a realistic case ions are also involved in the plasma motion. We can consider ion-acoustic waves as an example of ion oscillations in plasma. The dispersion relation for these waves reads~\cite{LifPit89},
\begin{equation}
    \omega(\mathbf{q}) = \omega_i
    \frac{\lambda_\mathrm{D}|\mathbf{q}|}{\sqrt{1 + \lambda_\mathrm{D}^2|\mathbf{q}|^2}}
\end{equation}
where $\omega_{i} = \omega_{e}\sqrt{m_{e}/m_{i}}$ is the Langmuir frequency for ions, $\lambda_\mathrm{D} = \sqrt{k_{\mathrm{B}} T_e / 4 \pi e^2 n_e^{(0)}}$ is the Debye length, $T_e$ is the electron temperature, and  $k_{\mathrm{B}}$ is the Boltzmann constant. Considering rather short waves with $|\mathbf{q}| \gg 1/\lambda_\mathrm{D}$ we get that $\omega \approx \omega_i$. Note that the damping of such waves is small~\cite{LifPit89}.

The oscillatory motion of ions can be also excited when various electron-ion and electron-electron nonlinearities are taken into account (see, e.g., Refs.~\cite{Zak72,SkoHaa80,DavYakZal05}). It makes Eq.~(\ref{eq:classeqmot}) highly nonlinear.
As a rule, these nonlinearities reduce
the frequency of electron oscillations. In this case we can assume that
electrons still perform oscillatory motion but with another frequency
$\omega_{0}<\omega_{e}$. Note that, in some cases, quantum effects can be more strongly marked for ions rather than for electrons~\cite{Abr61}.

Thus we will take that the motion of a charged
particle in plasma is governed by the potential $V(\mathbf{r})=m\omega_{0}^{2}r^{2}/2$,
where $m$ stays for the electron or ion mass and $\omega_{0}$ is
the oscillations frequency of charged particles. As we mentioned above, oscillations of ions should be treated as the collective effect.

Let us assume that the dynamics of charged particles in a quantum
plasmoid is also determined by the same potential $V(\mathbf{r})$.
Using this assumption, the construction of the ground state of this
plasma structure will be made in Sec.~\ref{sec:QUANTEL}. Of course,
to have a correlated many particles system, the typical size of
such an object should be quite small. We shall obtain some estimates
of the length scale of these plasmoids in Sec~\ref{sec:APPL}. Note
that, in Ref.~\cite{Daw59}, the cases of planar, cylindrical, and
spherical plasma oscillations were studied. In our work we will be
mainly concentrated on the description of spherically symmetric quantum
plasmoids.

\section{Quantization of the charged particles motion\label{sec:QUANTEL}}

To describe the quantum dynamics of a particle in plasma we choose
the Schr\"{o}dinger picture, i.e. the operators of observables will be
time independent. If we study the stationary states of particles,
with the total wave function $\sim\exp\left(-\mathrm{i}\tfrac{Et}{\hbar}\right)\psi$,
where $E$ is the particle energy, then the Schr\"{o}dinger equation for
the coordinate wave function $\psi$ has the form,
\begin{equation}\label{eq:Scheqelgen}
  E\psi = \hat{H}_{\mathrm{fq}}\psi,
  \quad
  \hat{H}_{\mathrm{fq}} = -\frac{\hbar^{2}}{2m}\nabla^{2}+\frac{m\omega_{0}^{2}r^{2}}{2},
\end{equation}
where we take into account that particles in a plasmoid move in the
potential $V(\mathbf{r})$, defined in Sec.~\ref{sec:MODQUANTPLAS}.

We shall be interested in the spherically symmetric solutions of Eq.~\eqref{eq:Scheqelgen}. Using the results of Ref.~\cite{CohDiuLal77} we can express the properly normalized total
wave function in the following form:
\begin{align}\label{eq:solpsi}
  \psi_{n\sigma}(r) = & \frac{1}{\sqrt{4\pi(2n+1)!}}
  \frac{(-1)^{n}}{2^{n}}
  \notag
  \\
  \times &
  \left(
    \frac{m^{3}\omega_{0}^{3}}{\pi\hbar^{3}}
  \right)^{1/4}
  \exp
  \left[
    -\frac{m\omega_{0}}{2\hbar}r^{2}
  \right]
  \notag
  \\
  \times &
  \frac{H_{2n+1}
  \left(
    \sqrt{\frac{m\omega_{0}}{\hbar}}r
  \right)
  }
  {\sqrt{\frac{m\omega_{0}}{\hbar}}r}
  \chi_{\sigma},
\end{align}
where $H_{n}(z)$ is the Hermite polynomial, $\chi_{\sigma}$ is the
spin wave function, and $\sigma = \pm 1$ is the spin variable. Quantum states, described by the wave functions~\eqref{eq:solpsi}, have the energies,
\begin{equation}\label{eq:elenn}
  E_{n} = \hbar\omega_{0}
  \left(
    2n+\frac{3}{2}
  \right),
\end{equation}
where $n=0,1,2,\dotsc$ is the radial quantum number.

To secondly quantize the system, we introduce the operator valued
wave functions,
\begin{equation}\label{eq:elwfsq}
 \hat{\psi}(r) = \sum_{n\sigma}\hat{a}_{n\sigma}\psi_{n\sigma}(r),
\end{equation}
where $\hat{a}_{n\sigma}$ is the annihilation operator. One can also
write down the analogous expression for the Hermitian conjugate wave
function which will contain the creation operator $\hat{a}_{n\sigma}^{\dagger}$.
We shall suppose that ions are half integer spin particles, i.e. they
are fermions like electrons. Thus, the operators $\hat{a}_{n\sigma}$
and $\hat{a}_{n\sigma}^{\dagger}$ satisfy the canonical anticommutation
relation $\left[\hat{a}_{n\sigma},\hat{a}_{n'\sigma'}^{\dagger}\right]_{+} = \delta_{nn'}\delta_{\sigma\sigma'}$
for both electrons and ions.

Using Eqs.~(\ref{eq:Scheqelgen}) and~(\ref{eq:solpsi}) we can
express the total energy of the system of charged particles as
\begin{align}
  E_{\mathrm{cp}} = & \int\mathrm{d}^{3}\mathbf{r}\hat{\psi}^{\dagger}(r)\hat{H}_{\mathrm{fq}}\hat{\psi}(r)
  \notag
  \\
  & =
  \sum_{n\sigma}E_{n}\hat{a}_{n\sigma}^{\dagger}\hat{a}_{n\sigma}.
\end{align}
Finally we replace the total energy
\begin{equation}\label{eq:Helsq}
  E_{\mathrm{cp}} \to \hat{H}_{\mathrm{cp}} = \sum_{n\sigma}E_{n}\hat{a}_{n\sigma}^{\dagger}\hat{a}_{n\sigma},
\end{equation}
with the secondly quantized Hamiltonian.

Now we can calculate the energy, $E_{0}$, of the ground state of
the system. If the total number of particles, $N$, is even, using
Eq.~(\ref{eq:Helsq}) we get for $E_{0}$,
\begin{align}\label{eq:E0Neven}
  E_{0} = & 2\sum_{n=0}^{n_{\mathrm{F}}}\hbar\omega_{0}
  \left(
    2n+\frac{3}{2}
  \right)
  \notag
  \\
  & =
  2\hbar\omega_{0}
  \left(
    n_{\mathrm{F}}^{2}+\frac{5}{2}n_{\mathrm{F}}+\frac{3}{2}
  \right).
\end{align}
If $N$ is odd, we analogously obtain that
\begin{align}\label{eq:E0Nodd}
  E_{0} = & 2\sum_{n=0}^{n_{\mathrm{F}}-1}\hbar\omega_{0}
  \left(
    2n+\frac{3}{2}
  \right) +
  \hbar\omega_{0}
  \left(
    2n_{\mathrm{F}}+\frac{3}{2}
  \right)
  \notag
  \\
  & =
  2\hbar\omega_{0}
  \left(
    n_{\mathrm{F}}^{2}+\frac{3}{2}n_{\mathrm{F}}+\frac{3}{4}
  \right),
\end{align}
Note that, if $N\to\infty$, both Eqs.~(\ref{eq:E0Neven}) and~(\ref{eq:E0Nodd})
approach the same limit $E_{0}\to2\hbar\omega_{0}n_{\mathrm{F}}^{2}$.
Here $n_{\mathrm{F}}$ stays for the maximal number of an occupied
energy level. We can call it the \emph{Fermi number}, in the analogy
to the Fermi momentum.

Let us define the effective size of a plasmoid, $R$, as the position
of the last maximum of the function $|\psi_{n_{\mathrm{F}}}(r)|^{2}$.
If $n_{\mathrm{F}}$ is great, this maximum is approximately achieved
at the classical turn point~\cite{Blo61}, i.e. when
$\hbar\omega_{0}\left(2n_{\mathrm{F}}+3/2\right) = m\omega_{0}^{2}R^{2}/2$.
Accounting for this fact, we get that
\begin{equation}\label{eq:kF}
  n_{\mathrm{F}} \approx \frac{m\omega_{0}R^{2}}{4\hbar}.
\end{equation}
Using Eq.~(\ref{eq:kF}) we can express the ground state energy in
terms of the plasmoid radius as $E_{0}\approx m^{2}\omega_{0}^{3}R^{4}/8\hbar$.

Note that our plasmoid model is based on the linear approximation
of noninteracting particles. It corresponds to the Hartree-Fock approximation. As we mentioned in Sec.~\ref{sec:MODQUANTPLAS},
if we account for various (classical or quantum) electron-ion and
electron-electron interactions, it will lead to the nonlinear terms
in the equation of motion~(\ref{eq:classeqmot}) of a particle and
impose a certain length scale in the system (see, e.g., Refs.~\cite{SkoHaa80,DavYakZal05}). One the examples of such
interaction is later discussed in Sec.~\ref{sec:INTELACOU}.

The ground state of a spherical plasmoid was constructed under the
assumption of the negligible particle temperature. The temperature
corrections will wash out the lowest energy levels with $E_{n}\lesssim k_{\mathrm{B}}T$,
where $T$ is the
electron or ion temperature. Taking into account the fact that particles
possessing the lowest energy levels are concentrated near the center
of a plasmoid, it means that plasma oscillations are the most intensive
in a spherical shell. The internal radius of this shell is determined
by the thermal effects and the external one by the nonlinear interactions
between particles. Such kind of the spherical plasmoid structure is
in agreement with the numerical simulations made in Ref.~\cite{DavYakZal05}

\section{Quantization of the acoustic field\label{sec:QUANTPH}}

As we mentioned in Sec.~\ref{sec:QUANTEL}, the temperature in plasma
should be not quite high for the existence of a spherical quantum
plasma structure. It means that, along with the charged particles,
there should be a quite significant fraction of neutral atoms or molecules
in plasma. Plasma oscillations will inevitably result in collisions
with neutral atoms leading to the generation of acoustic waves. In
this Section we shall quantize acoustic waves inside a spherical plasmoid.

The hydrodynamics equations, which govern the acoustic waves propagation,
have the form,
\begin{align}\label{eq:hydrodyngen}
  \frac{\partial n_{n}}{\partial t} +
  \nabla(n_{n}\mathbf{v}_{n}) & =0,\nonumber \\
  \frac{\partial\mathbf{v}_{n}}{\partial t}+(\mathbf{v}_{n}\nabla)\mathbf{v}_{n} +
  \frac{1}{n_{n}m_{n}}\nabla p_{n} & =0,
\end{align}
where $n_{n}$ is the number density of neutral particles, $\mathbf{v}_{n}$
is the neutral gas velocity, $p_{n}(n_{n})$ is the pressure of the
gas, and $m_{n}$ is the neutral particle mass.

Let us expand the
parameters $n_{n}$, $\mathbf{v}_{n}$, and $p_{n}$ as
\begin{align}\label{eq:hydrodyndecom}
  n_{n}= & n_{n}^{(0)}+n_{1} + \dotsb,\nonumber \\
  \mathbf{v}_{n}= & \mathbf{v}_{1} + \dotsb,\nonumber \\
  p_{n}= & p_{n}^{(0)} +
  \left(
    \frac{\partial p_{n}}{\partial n_{n}}
  \right)_{0}
  (n_{n}-n_{n}^{(0)})+\dotsb,
\end{align}
where $n_{n}^{(0)}$ and $p_{n}^{(0)}$ are the equilibrium values
of the number density and the pressure as well as $n_{1}$ and $\mathbf{v}_{1}$
are the small perturbations of the number density and the velocity. Using Eqs.~\eqref{eq:hydrodyngen} and~\eqref{eq:hydrodyndecom} we get two independent wave equations for $n_{1}$ and $\mathbf{v}_{1}$,
%
%
\begin{align}
  \frac{\partial^{2}\mathbf{v}_{1}}{\partial t^{2}} -
  c_{\mathrm{s}}^{2}\nabla(\nabla\cdot\mathbf{v}_{1}) & =0,\label{eq:wev1}\\
  \frac{\partial^{2}n_{1}}{\partial t^{2}} -
  c_{\mathrm{s}}^{2}\nabla^{2}n_{1} & =0,\label{eq:wen1}
\end{align}
where
\begin{equation}\label{soundv}
  c_{s} = \frac{1}{m_{n}^{1/2}}\sqrt{
  \left(
    \frac{\partial p_{n}}{\partial n_{n}}
  \right)_{0}},
\end{equation}
is the sound velocity. Although we linearize Eq.~\eqref{eq:hydrodyndecom}, the influence of charged particles on the neutral component of plasma can be accounted for by renormalizing the sound velocity~\eqref{soundv}, which is regarded as a phenomenological parameter.

The spherically symmetric solution of Eq.~(\ref{eq:wen1}) has the
form,
\begin{equation}\label{eq:phonwf}
  n_{1}(\mathbf{r},t) = e^{-\mathrm{i}\omega_{k}t}f_{k}(r),
  \quad
  f_{k}(r)=\frac{\sin kr}{kr},
\end{equation}
where $\omega_{k}=c_{s}k$. Note that the coordinate function obeys
the following identities:
\begin{align}
  \int\frac{\mathrm{d}^{3}\mathbf{r}}{(2\pi)^{3}}f_{k}(r)f_{p}(r) = & \delta^{3}(\mathbf{k}-\mathbf{\mathbf{p}}),
  \notag
  \\
  \int\frac{\mathrm{d}^{3}\mathbf{k}}{(2\pi)^{3}}f_{k}(r)f_{k}(r') = & \delta^{3}(\mathbf{r}-\mathbf{r}'),
\end{align}
where delta-functions should be understood as $\delta^{3}(\mathbf{k}-\mathbf{\mathbf{k}}')=\delta(k-k')/4\pi k^{2}$.

Following Ref.~\cite{ZakKuz97}, we introduce the potential of the velocity
as $\mathbf{v}_{1}=\nabla\varphi_{1}$. Using Eqs.~(\ref{eq:wev1})-(\ref{eq:phonwf})
we obtain the general expressions for $n_{1}$ and $\varphi_{1}$
as
\begin{align}\label{eq:n1phi1bbp}
  n_{1}(r,t)= & \int\frac{\mathrm{d}^{3}\mathbf{k}}{(2\pi)^{3/2}}
  \left(
    \frac{\hbar n_{n}^{(0)}}{2m_{n}\omega_{k}}
  \right)^{1/2}
  \notag
  \\
  & \times
  kf_{k}
  \left(
    b_{k}e^{-\mathrm{i}\omega_{k}t}+b_{k}^{\dagger}e^{\mathrm{i}\omega_{k}t}
  \right),\nonumber \\
  \varphi_{1}(r,t)= & \int\frac{\mathrm{d}^{3}\mathbf{k}}{(2\pi)^{3/2}}
  \left(
    \frac{\hbar}{2m_{n}n_{n}^{(0)}\omega_{k}}
  \right)^{1/2}
  \notag
  \\
  & \times
  c_{s}f_{k}
  \left(
    -\mathrm{i}b_{k}e^{-\mathrm{i}\omega_{k}t}+\mathrm{i}b_{k}^{\dagger}e^{\mathrm{i}\omega_{k}t}
  \right),
\end{align}
where $b_{k}$ and $b_{k}^{\dagger}$ are the Fourier coefficients.

The total energy of the acoustic field has the form,
\begin{equation}
  E_{\mathrm{ph}} =
  \frac{1}{2}\int\mathrm{d}^{3}\mathbf{r}\ m_{n}
  \left[
    n_{n}^{(0)}\nabla\varphi_{1}^{2}+\frac{c_{s}^{2}}{n_{n}^{(0)}}n_{1}^{2}
  \right].
\end{equation}
Now, if we replace $b_{k}$ and $b_{k}^{\dagger}$ in Eq.~(\ref{eq:n1phi1bbp})
by the annihilation and the creation operators, $\hat{b}_{k}$ and
$\hat{b}_{k}^{\dagger}$, we can express $E_{\mathrm{ph}}$, which
should be regarded as the secondly quantized Hamiltonian of the phonon
field, in the following way:
\begin{align}\label{eq:Hphsq}
  E_{\mathrm{ph}} \to \hat{H}_{\mathrm{ph}} = &
  \int\mathrm{d}^{3}\mathbf{k}\ \hbar\omega_{k}\ \hat{b}_{k}^{\dagger}\hat{b}_{k}
  \notag
  \\
  & +
  \text{divergent terms}.
\end{align}
To derive Eq.~(\ref{eq:Hphsq}) we suggest that the operators $\hat{b}_{k}$
and $\hat{b}_{k}^{\dagger}$ obey the canonical commutation relation
for bosonic operators, $[\hat{b}_{k},\hat{b}_{k'}^{\dagger}]_{-}=\delta^{3}(\mathbf{k}-\mathbf{k}')$.
Note the ``divergent terms'' in Eq.~(\ref{eq:Hphsq}), containing
$\delta(0)$, can be removed by the normal ordering of operators.

Note that, in this Section, we used the Heisenberg picture where operators
turn out to be time dependent, cf. Eq.~(\ref{eq:n1phi1bbp}). Using results of Ref.~\cite{AbrGorDzy65} one can express these operators in the Schr\"{o}dinger picture, which will be used in Sec.~\ref{sec:INTELACOU}. Such a transformation
is equivalent to setting $t=0$ in Eq.~(\ref{eq:n1phi1bbp}).


\section{Interaction between charged particles and acoustic waves\label{sec:INTELACOU}}

We have already mentioned in Sec.~\ref{sec:QUANTPH} that charged
particles can interact with neutral atoms, which are present in the
system. Thus charged particles should also interact with acoustic
waves, which are generated by spherically symmetric plasma oscillations.
In this Section we derive the secondly quantized Hamiltonian which
describes this interaction. Then we exclude the acoustic part of the
total Hamiltonian and reduce it to the nonlinear charged particles
Hamiltonian.

We suggest that charged particles scatter off the density perturbations
of neutral atoms or molecules. Thus the energy of the interaction
of a charged particle with an acoustic field has the form~\cite{VlaYak78},
\begin{equation}\label{eq:intenelph}
  V(\mathbf{r}) = \int\mathrm{d}^{3}\mathbf{r}'\ K(\mathbf{r}-\mathbf{r}')n_{1}(\mathbf{r}'),
\end{equation}
where $K(\mathbf{r}-\mathbf{r}')$ is the energy of the interaction between
a charged particle placed at $\mathbf{r}$ and a neutral particle which is
at $\mathbf{r}'$. Using Eq.~(\ref{eq:intenelph}) we can derive
the secondly quantized Hamiltonian of the interaction between charged particles and acoustic
waves as
\begin{align}\label{eq:Helphrsq}
  \hat{H}_{\mathrm{cp-ph}} = & \int\mathrm{d}^{3}\mathbf{r}
  \hat{\psi}^{\dagger}(\mathbf{r})V(\mathbf{r})\hat{\psi}(\mathbf{r})
  \notag
  \\
  & =
  K_{0} \int\mathrm{d}^{3}\mathbf{r}
  \hat{\psi}^{\dagger}(\mathbf{r})\hat{\psi}(\mathbf{r})\hat{n}_{1}(\mathbf{r}),
\end{align}
where $\hat{\psi}$ and $\hat{n}_{1}$ are given in Eqs.~(\ref{eq:elwfsq})
and~(\ref{eq:n1phi1bbp}) respectively. To derive Eq.~(\ref{eq:Helphrsq})
we suggest that $K(\mathbf{r}-\mathbf{r}')=K_{0}\delta^{3}(\mathbf{r}-\mathbf{r}')$.
This approximation corresponds to a contact interaction between
charged and neutral particles.

Using Eqs.~(\ref{eq:elwfsq}) and~(\ref{eq:n1phi1bbp}) we can cast
$\hat{H}_{\mathrm{el-ph}}$ to the form,
\begin{align}\label{eq:Helphsq}
  \hat{H}_{\mathrm{cp-ph}} = &
  \int\frac{\mathrm{d}^{3}\mathbf{k}}{(2\pi)^{3/2}}
  \sum_{ns\sigma}D_{ns}(k)
  \notag
  \\
  & \times
  \hat{a}_{n\sigma}^{\dagger}\hat{a}_{s\sigma}
  \left(
    \hat{b}_{k}+\hat{b}_{k}^{\dagger}
  \right),
\end{align}
where
\begin{align}
  D_{ns}(k) = & K_{0}k
  \left(
    \frac{\hbar n_{n}^{(0)}}{2m_{n}\omega_{k}}
  \right)^{1/2}
  \notag
  \\
  & \times
  \int\mathrm{d}^{3}\mathbf{r}\ \psi_{n}(r)\psi_{s}(r)f_{k}(r),
\end{align}
is the matrix element. With help of Eqs.~(\ref{eq:solpsi}) and~(\ref{eq:phonwf})
we can compute $D_{ns}(k)$ at big $n$ and $s$ in the explicit form,
\begin{align}
  D_{ns}(k)\approx & \frac{K_{0}}{4(ns)^{1/4}}
  \left(
    \frac{n_{n}^{(0)}m\omega_{0}}{2m_{n}\omega_{k}}
  \right)^{1/2}
  \notag
  \\
  & \times
  [\mathrm{sign}(2\sqrt{n}-2\sqrt{s}+\xi)
  \notag
  \\
  & +
  \mathrm{sign}(2\sqrt{n}+2\sqrt{s}-\xi)
  \notag
  \\
  & -
  \mathrm{sign}(2\sqrt{n}-2\sqrt{s}-\xi)-1],\label{eq:Dnmapr}
\end{align}
where $\xi=k\sqrt{\hbar/m\omega_{0}}$.%

To exclude the acoustic degrees of freedom we make the canonical transformation
of the Hamiltonian in Eq.~(\ref{eq:Helphsq}),
\begin{align}
  \hat{H}_{\mathrm{cp-ph}} & \to e^{-\hat{S}}\hat{H}_{\mathrm{cp-ph}}e^{\hat{S}},\nonumber \\
  \hat{S}= & \int\frac{\mathrm{d}^{3}\mathbf{k}}{(2\pi)^{3/2}}
  \sum_{ns\sigma}D_{ns}(k)\hat{a}_{n\sigma}^{\dagger}\hat{a}_{s\sigma}
  \notag
  \\
  & \times
  \bigg(
    \frac{\hat{b}_{k}}{E_{s}-E_{n}-\hbar\omega_{k}}
    \notag
    \\
    & +
    \frac{\hat{b}_{k}^{\dagger}}{E_{s}-E_{n}+\hbar\omega_{k}}
  \bigg).
\end{align}
It can be seen that, after this transformation, the total Hamiltonian,
$\hat{H}=\hat{H}_{0}+\hat{H}_{\mathrm{cp-ph}}$, transforms into $\hat{H}\to\hat{H}_{0}+\tfrac{1}{2}[\hat{H}_{\mathrm{cp-ph}};\hat{S}]$,
where $\hat{H}_{0}=\hat{H}_{\mathrm{cp}}+\hat{H}_{\mathrm{ph}}$.
Averaging over the acoustic ground state and assuming that there are
no external phonons, we get the total Hamiltonian in the form,
\begin{align}\label{eq:Htotgen}
  \hat{H} = & \sum_{n\sigma} E_{n} \hat{a}_{n\sigma}^{\dagger} \hat{a}_{n\sigma}
  \notag
  \\
  & +
  \int\frac{\mathrm{d}^{3}\mathbf{k}}{(2\pi)^{3}}\sum_{nn'ss'\sigma\sigma'}D_{ns}(k)D_{n's'}(k)
  \notag
  \\
  & \times
  \hat{a}_{n\sigma}^{\dagger}\hat{a}_{n'\sigma'}^{\dagger}\hat{a}_{s'\sigma'}\hat{a}_{s\sigma}
  \notag
  \\
  & \times
  \frac{\hbar\omega_{k}}{(E_{s}-E_{n})^{2}-(\hbar\omega_{k})^{2}}.
\end{align}

Note that the Hamiltonian~(\ref{eq:Htotgen}) contains both the free
charged particles term and the interaction between charged particles
owing to the virtual acoustic wave exchange. Let us examine when this
effective interaction can be attractive. We remind that charged particles
in a plasmoid oscillate with a rather high frequency $\sim\omega_{0}.$
In collisions with neutral atoms these charged particles will generate
acoustic waves with the typical frequency $\sim\omega_{0}$~\cite{Dvo12}.
It means that the phonon energy in Eq.~(\ref{eq:Htotgen}) is $\hbar\omega_{k}\sim\hbar\omega_{0}$.
Using Eq.~(\ref{eq:elenn}) we get that $E_{n}\sim\hbar\omega_{0}n$.
It means that the effective interaction can be attractive only if
$E_{s}=E_{n}$ or $s=n$. Thus there is an attraction between a pair of charged particles which are at the same energy level.

\section{Bound states of charged particles\label{sec:DIAGHAM}}

In the previous Section we have found that the exchange of a virtual
acoustic wave can result in the attractive interaction between oscillating
charged particles in plasma. In this Section we study this process
in details and show that under certain conditions charged particles
can form pairs.

Before we proceed it should be noted that charged particles with parallel
spins cannot occupy the same energy level, because of the Pauli principle. We remind that ions in our system are
supposed to be fermions. Thus we should exclude the contribution of
charged particles with parallel spins from the effective interaction~(\ref{eq:Htotgen}).
It mean that the summation in the nonlinear term should be made over the opposite spin indexes
$\sigma'=-\sigma$.

We should also mention that up to now we studied the case of constant
number of particles. To avoid this restriction, we use the standard
technique of shifting the energy levels $E_{n}\to e_{n}=E_{n}-\mu$,
where $\mu$ is the chemical potential of the system. Now the number
of particles can change, but we should calculate the chemical potential.

Finally, taking into account all these comments we can transform the
Hamiltonian~(\ref{eq:Htotgen}) to the following form:
\begin{align}\label{eq:Htotaps}
  \hat{H} = &
  \sum_{n\sigma}
  e_{n}\hat{a}_{n\sigma}^{\dagger}\hat{a}_{n\sigma}
  \notag
  \\
  & -
  \sum_{nn'\sigma}
  F_{nn'}\hat{a}_{n\sigma}^{\dagger}\hat{a}_{n',-\sigma}^{\dagger}\hat{a}_{n',-\sigma}\hat{a}_{n\sigma},
\end{align}
where the new matrix element can be explicitly calculated on the basis
of Eq.~(\ref{eq:Dnmapr}) as%
\begin{align}\label{eq:atrpot}
  F_{nn'} = & V_{0}\frac{
    \sqrt{n}+\sqrt{n'}-|\sqrt{n}-\sqrt{n'}|
  }{\sqrt{nn'}},
  \notag
  \\
  V_{0} = & \frac{K_{0}^{2}n_{n}^{(0)}}{8\pi^{2}m_{n}c_{s}^{2}}
  \left(
    \frac{m\omega_{0}}{\hbar}
  \right)^{3/2}.
\end{align}
It is worth mentioning that $F_{nn'}>0$ for any $n$ and $n'$. It
means that the effective interaction described by the Hamiltonian~(\ref{eq:Htotaps})
is really attractive.

To diagonalize the Hamiltonian~(\ref{eq:Htotaps}) we introduce the
new operators $\hat{A}_{0n}$ and $\hat{A}_{1n}$ by means of the
Bogolyubov transformation,
\begin{align}\label{eq:newoper}
  \hat{a}{}_{n+} = & u_{n} \hat{A}_{0n} + v_{n}\hat{A}_{1n}^{\dagger},
  \notag
  \\
  \hat{a}_{n-} = & u_{n}\hat{A}_{1n} - v_{n}\hat{A}_{0n}^{\dagger},
\end{align}
where the real coefficients $u_{k}$ and $v_{k}$ satisfy the relation
$u_{n}^{2}+v_{n}^{2}=1$. Note that the new operators obey the canonical
anticommutation relations for fermion operators. Substituting these
operators in Eq.~(\ref{eq:Htotaps}) we get the Hamiltonian in the
form,
\begin{equation}\label{eq:Hscraw}
  \hat{H} = E'_{0}+\hat{H}_{2}+\hat{H}'_{2}+\dotsc,
\end{equation}
where
\begin{equation}\label{eq:groundstsc}
  E'_{0} = 2\sum_{n}v_{n}^{2}
  \left[
    e_{n}-F_{nn}u_{n}^{2}
  \right],
\end{equation}
is the ground energy,
\begin{align}\label{eq:diagint}
  \hat{H}_{2} = & \sum_{n}
  \left[
    e_{n}
    \left(
      u_{n}^{2}-v_{n}^{2}
    \right) + 4F_{nn}u_{n}^{2}v_{n}^{2}
  \right]
  \notag
  \\
  & \times
  \left(
    \hat{A}_{1n}^{\dagger}\hat{A}_{1n}+\hat{A}_{0n}^{\dagger}\hat{A}_{0n}
  \right),
\end{align}
is the energy of the new quasiparticles, and
\begin{align}\label{eq:nondaigint}
  \hat{H}'_{2} = & 2\sum_{n}u_{n}v_{n}
  \left[
    e_{n}-F_{nn}
    \left(
      u_{n}^{2}-v_{n}^{2}
    \right)
  \right]
  \notag
  \\
  & \times
  \left(
    \hat{A}_{1n}\hat{A}_{0n}+\hat{A}_{0n}^{\dagger}\hat{A}_{1n}^{\dagger}
  \right),
\end{align}
is the nondiagonal part of the quasiparticles interaction. Note that
in Eq.~(\ref{eq:Hscraw}) we omit terms higher than quadratic ones.

The nondiagonal interaction~(\ref{eq:nondaigint}) can be vanishing
in two cases. Firstly, when either $u_{n}=1$ and $v_{n}=0$ or $u_{n}=0$
and $v_{n}=1$. It corresponds to the trivial solution, which is equivalent
to the transition to the particles-holes representation. Note that
the ground state energy~(\ref{eq:groundstsc}) is not negative for
this trivial solution: $E'_{0}\geq0$.

The second situation is implemented if the attraction between charged
particles is rather strong, i.e. when
\begin{equation}\label{eq:boundcond}
  F_{nn}>|e_{n}|.
\end{equation}
In this case we can choose the coefficients $u_{k}$ and $v_{k}$
as
\begin{align}\label{eq:unvu}
  u_{n}^{2} = & \frac{1}{2}
  \left(
    1+\frac{e_{n}}{F_{nn}}
  \right),
  \notag
  \\
  v_{n}^{2} = & \frac{1}{2}
  \left(
    1-\frac{e_{n}}{F_{nn}}
  \right).
\end{align}
The ground state energy~(\ref{eq:groundstsc}) now becomes negative,
\begin{equation}
  E_{0} = -\sum_{n}\frac{(F_{nn}-e{}_{n})^{2}}{2F_{nn}}<0.
\end{equation}
It means that the state of the system, corresponding to the new quasiparticles,
is more favorable energetically than in the trivial case discussed
above. The diagonal part of the Hamiltonian~(\ref{eq:diagint}) has
the form,
\begin{equation}\label{eq:diagintqp}
  \hat{H}_{2} = \sum_{n}F_{nn}
  \left(
    \hat{A}_{1n}^{\dagger}\hat{A}_{1n}+\hat{A}_{0n}^{\dagger}\hat{A}_{0n}
  \right).
\end{equation}
Using Eqs.\eqref{eq:atrpot} and~(\ref{eq:diagintqp}) one can find the energy of quasiparticles
$E'_{n}=2V_{0}/\sqrt{n}$.

The chemical potential of the system can be calculated using the expression
$N=\langle\Phi'_{0}|\hat{N}|\Phi'_{0}\rangle$, where $|\Phi'_{0}\rangle$
is the wave function of the ground state satisfying the conditions
\begin{equation}\label{eq:vaccond}
  \hat{A}_{0n}|\Phi'_{0}\rangle=0,
  \quad
  \text{and}
  \quad
  \hat{A}_{1n}|\Phi'_{0}\rangle=0,
\end{equation}
and
\begin{equation}\label{eq:numpartoper}
  \hat{N}=\sum_{n\sigma}\hat{a}_{n\sigma}^{\dagger}\hat{a}{}_{n\sigma},
\end{equation}
is the number of particles operator.

Using Eqs.~(\ref{eq:newoper}), (\ref{eq:unvu}), (\ref{eq:vaccond}),
and~(\ref{eq:numpartoper}) we can express the number of particles
as
\begin{equation}\label{eq:numpartcond}
  N=2\sum_{n}v_{n}^{2}=\sum_{n<n{}_{\mathrm{F}}'}
  \left(
    1-\frac{e_{n}}{F_{nn}}
  \right),
\end{equation}
where the bound of the new state in the energy space, $n_{\mathrm{F}}'$,
is defined by Eq.~(\ref{eq:boundcond}). The quantity $n'_{\mathrm{F}}$
is analogous to the Fermi number introduced in Sec.~\ref{sec:QUANTEL}.
Basing on Eqs.~(\ref{eq:elenn}) and~(\ref{eq:numpartcond}) we
get the system of equations,
\begin{align}\label{eq:NnFmu}
  N= & \frac{n'_{\mathrm{F}}}{3}+\frac{4}{15}n_{\mathrm{F}}^{\prime5/2}\frac{\hbar\omega_{0}}{V_{0}},\nonumber \\
  \mu= & \hbar\omega_{0}
  \left(
    2n'_{\mathrm{F}}+\frac{3}{2}
  \right)-\frac{2V_{0}}{\sqrt{n'_{\mathrm{F}}}},
\end{align}
which define the quantities $n'_{\mathrm{F}}$ and $\mu$ as functions
of $N$. To derive Eq.~(\ref{eq:NnFmu}) we use the approximate identity,
\begin{equation}
 \sum_{n<n{}_{\mathrm{F}}'}n^{\alpha} \approx \int_{0}^{n{}_{\mathrm{F}}'}n^{\alpha}\mathrm{d}n=\frac{(n{}_{\mathrm{F}}')^{\alpha+1}}{\alpha+1},
\end{equation}
which is valid at big $n{}_{\mathrm{F}}'$.

The analysis of this Section shows that the exchange of a virtual acoustic
wave between oscillating charged particles in a quantum spherical
plasmoid results in the formation of pairs of these particles. There is a
pairing of particles with oppositely directed spins. These particles
should be at the same energy level. Note that this new state of plasma
has less energy of the ground state compared to the situation of unpaired
particles. This process is analogous to the formation of Cooper pairs of electrons
in a metal. We remind that, if the temperature of a metal is significantly
low, two electrons can form a bound state owing to the exchange of
a virtual phonon, which is a quanta of the vibration of a crystal
lattice of a metal. This process underlies the phenomenon of superconductivity.

Note that besides the oppositely directed spins, two electrons in
a metal should have opposite momenta~\cite{Mad78}. In our case
the situation is analogous. Indeed, at big radial quantum numbers, $n\gg1$,
the asymptotic expansion of the charged particles wave function~(\ref{eq:solpsi})
can be expressed as
\begin{align}\label{eq:elwfexp}
  \psi_{n}(r) \sim & \frac{1}{r}
  \sin
  \left(
    2\sqrt{\frac{m\omega_{0}n}{\hbar}}r
  \right)
  \notag
  \\
  & =
  \frac{1}{2\mathrm{i}r}
  \bigg[
    \exp
    \left(
      2\mathrm{i}\sqrt{\frac{m\omega_{0}n}{\hbar}}r
    \right)
    \notag
    \\
    & -
    \exp
    \left(
      -2\mathrm{i}\sqrt{\frac{m\omega_{0}n}{\hbar}}r
    \right)
  \bigg].
\end{align}
Eq.~(\ref{eq:elwfexp}) means that any steady state of a charged
particle in a spherical plasmoid is a superposition of converging
and divergent spherical running waves. The pairing of two charged
particles with opposite spins happens between the states corresponding
to the running waves with opposite momenta.%

Let us demonstrate by means of the explicit calculation that the formation
of a bound state of two charged particles is possible in our system.
Suppose that there is the state of two charged particles having opposite
spins,
\begin{equation}\label{eq:wfcooppair}
  |\Psi\rangle = \sum_{n}c_{n}
  \hat{a}_{n,+}^{\dagger}
  \hat{a}_{n,-}^{\dagger}
  |\Phi_{0}\rangle,
\end{equation}
where $c_{n}$ are the expansion coefficients and $|\Phi_{0}\rangle$
is the ground state corresponding to the filled lowest energy states:
$\hat{a}_{n\sigma}|\Phi_{0}\rangle=0$, cf. Sec.~\ref{sec:QUANTEL}. Using
the Hamiltonian~(\ref{eq:Htotaps}) we can show that the energy corresponding
to the state~(\ref{eq:wfcooppair}) is
\begin{equation}\label{eq:energycooppair}
  E_{\Psi} = \langle\Psi|\hat{H}|\Psi\rangle=2\sum_{n}
  \left(
    E_{n}-F_{nn}
  \right)
  |c_{n}|^{2},
\end{equation}
where $E_{n}$ is given by Eq.~(\ref{eq:elenn}). Thus the energy
of the state~(\ref{eq:wfcooppair}) is less than the sum of energies
of two non-interacting charged particles, with $F_{nn}$ being the
binding energy.

Analyzing Eq.~(\ref{eq:energycooppair}) it is interesting to mention
that, in contrast to the formation of Cooper pairs in metal, where
bound states are formed near the Fermi sphere surface, in our case
the pairing of particles happens at all the energy levels. This
fact can be explained by the very convenient choice of the ground
state wave functions~(\ref{eq:solpsi}), which maximally account
for the dynamical features of the system (see also Sec.~\ref{sec:MODQUANTPLAS}).
Of course, we should remind that our analysis is valid for relatively
big radial quantum numbers $n$.

\section{Possible applications\label{sec:APPL}}

To analyze the possibility of the pairing of charged particles in a spherically
symmetric plasma structure we can use Eq.~(\ref{eq:energycooppair})
which gives one the energy of the bound state of two particles with
opposite spins. Note that this expression is valid just before the
pairing since it is based on the ground state $|\Phi_{0}\rangle$
rather than on $|\Phi'_{0}\rangle$, defined by Eq.~(\ref{eq:vaccond}).
The pairing may happen if $E_{n}<F_{nn}$. Using Eqs.~(\ref{eq:elenn})
and~(\ref{eq:atrpot}), one obtains that this constraint is equivalent to
\begin{equation}\label{eq:condcondn}
  n<
  \left(
    \frac{V_{0}}{\hbar\omega_{0}}
  \right)^{2/3}.
\end{equation}
Supposing that the Fermi number of a plasmoid before the pairing,
given in Eq.~(\ref{eq:kF}), is smaller than the constraint~(\ref{eq:condcondn}),
we get the following upper bound:
\begin{equation}\label{eq:condcondR}
  R<R_{\mathrm{cr}} =
  \left(
    \frac{K_{0}^{2}n_{n}^{(0)}}{\pi^{2}\hbar\omega_{0}m_{n}c_{s}^{2}}
  \right)^{1/3},
\end{equation}
on the radius of a plasma structure.

Note the constraint~(\ref{eq:condcondR}) is very conservative. It
means that charged particles on all energy levels can form bound states
simultaneously. On the contrary, we may require that only particles
at lower levels form pairs and then this process
spreads to higher energy states. Thus in a realistic case the upper
bound of plasmoid radius~(\ref{eq:condcondR}) may be significantly
changed towards its enhancement. However the analysis of the dynamics
of this phase transition requires a separate special study.

To evaluate the plasmoid radius one should define $K_{0}$, which
does not depend on the macroscopic plasma characteristics. We remind
that we approximated the potential $K(\mathbf{r})$ by a delta-function.
For this kind of potential, using the Born approximation, one can
calculate the total cross section of the charged particles scattering
off the neutral particles as $\sigma_s = m^{2}K_{0}^{2}/\pi\hbar^{4}$.
Note that in our approximation the cross section does not depend on
the particles energy, which may not be the case for some realistic potentials.

Let us first examine the possibility of pairing of charged particles
inside a spherical plasmoid in a very dense medium corresponding to
the inner crust of a neutron star (NS). The number density of neutral particles,
i.e. neutrons, at the bottom of the inner crust is $n_{n}^{(0)}\sim10^{38}\ \text{cm}^{-3}$~\cite{ShaTeu83}.
We will be mainly interested in the studies of the formation of pairs of protons.
The number density of protons in the NS crust strongly depends
on the equation of state of the NS matter. According to
Ref.~\cite{SheHorTie11} it can be about several per cent of the neutron
density. In our estimates we take that $n_{p}^{(0)}\sim10^{36}\ \text{cm}^{-3}$.
Thus we get that at such a density the Langmuir frequency for protons
is $\omega_{0}\sim10^{21}\ \text{s}^{-1}$. The sound velocity also
depends on the equation of state of the nuclear matter. Nevertheless
we may take it as $c_{s}\sim10^{9}\ \text{cm}\cdot\text{s}^{-1}$~\cite{Eps88},
which corresponds to the chosen density of the NS crust.

Protons in the NS crust are highly degenerate. Their Fermi
energy does not exceed several MeV. At such typical energies the cross
section of the proton-neutron scattering is approximately constant
and equals to $2\times10^{-23}\ \text{cm}^{2}$~\cite{Cha06}.
This fact justifies the use of the delta-function potential in Eq.~\eqref{eq:Helphrsq}.
Thus we can also evaluate $K_{0}$. Finally, using Eq.~(\ref{eq:condcondR})
we get that protons can form pairs in
a plasmoid with $R_{\mathrm{cr}}\approx4.6\times10^{-12}\ \text{cm}$.
The obtained value for the critical radius means that there can be
several hundreds of energy levels inside the plasmoid. This result
is in agreement with the fact that there can be a proton superconductivity
in the NS matter~\cite{YakLevShi99}.

It should be mentioned that besides uniform dense matter, containing neutrons, protons, and electrons, nuclei can be also present in the NS crust. The presence of nuclei can affect the process of the protons pairing in frames of our model. However, at the densities corresponding to the bottom of inner crust, used in our work, the fraction of nuclei is negligible~\cite{HaePotYak07}.

Note that the energy corresponding
to the zeroth level, $3\hbar\omega_{0}/2$, turns out to be of the
order of the Fermi energy for protons. Thus thermal effects will not
influence the plasmoid dynamics. We should also mention that for the pairing of protons inside the NS matter to happen, the proton-phonon interaction owing to the acoustic wave exchange should be dominant. It means that the electromagnetic interaction of protons should be screened. Supposing that electrons in the NS crust are ultrarelativistic, we get for the Debye length, $\lambda_\mathrm{D} \sim 10^{-13}\ \text{cm}$, that is much smaller than the plasmoid radius. Thus the electromagnetic interaction of protons can be omitted.

It should be noticed that the cross section of the electron-neutron
scattering is several orders of magnitude smaller than that of the
proton-neutron scattering~\cite{FroPap87}. Moreover the Langmuir frequency
for electrons is much higher than that for protons. It means that
the critical volume of a plasmoid becomes too small to contain enough number
of particles. Thus the electron superconductivity is unlikely to exist
in a spherical plasmoid inside NS.

Although one might expect that the pairing of charged particles
can happen only in plasmas with very high densities, which may be encountered
only in astrophysical media, we may discuss the situation when this
phenomenon occurs in terrestrial conditions. Let us discuss the case
when a plasmoid appears in liquid water with $n_{n}^{(0)}\sim10^{23}\ \text{cm}^{-3}$.
We suggest that the number density of singly ionized water ions is
$n_{i}^{(0)}\sim10^{21}\ \text{cm}^{-3}$. The Langmuir frequency
for water ions is $\omega_{0}\sim10^{13}\ \text{s}^{-1}$. The sound
speed in water is $c_{s}\sim10^{5}\ \text{cm}\cdot\text{s}^{-1}$.
Note that the total spin of a neutral water molecule is integer. Thus, if
we study a singly ionized water ion, its spin should be half integer,
i.e. it is a fermion.

Unfortunately, the scattering of water ions on neutral molecules of
water is not very well studied. Anyway we can assume that the total
cross section of such a scattering cannot be less than that for electron
scattering on water molecules which can be $\sim10^{-14}\ \text{cm}^{2}$
in the eV electron energy range~\cite{ItiMas05}. Thus using Eq.~(\ref{eq:condcondR})
we get that $R_{\mathrm{cr}}\approx4.6\times10^{-7}\ \text{cm}$.
Again we can see that there are several hundreds of energy levels
inside the plasmoid.

As in case of a plasmoid existing in the crust
of NS, the pairing of electrons in a spherical plasma
structure involving water ions is unlikely to happen because of the
very high frequency of electron oscillations. The fact that the exchange
of a virtual acoustic wave results in the cohesion of ions rather
than electrons was also noticed in Ref.~\cite{VlaYak78}.

The typical
energy of a stochastic motion of an ion is $k_{\mathrm{B}}T_{i}\sim10^{-14}\ \text{erg}$,
where $T_{i}\approx300\ \text{K}$ is the ion temperature. Using the
above estimate for the oscillations frequency, we get that the energy
of the zeroth energy level is also $\sim10^{-14}\ \text{erg}$. Thus
we can disdain the thermal effects in a spherical plasma structure
with water ions. We can also check that one can neglect the electromagnetic interaction between water ions. Taking the electron temperature of $\sim 10^3\ \text{K}$ (such a temperature corresponds to $10\%$ of the ionization potential of a hydrogen atom), we get that the Debye length is $\sim 10^{-8}\ \text{cm}$, which is much less than the plasmoid radius.

We have shown that the pairing of ions is possible in a plasmoid
which is a radial oscillation of charged particles in plasma. It is
also clear that ions play a subdominant role in plasma oscillations
because of their low mobility. We may put forward a hypothesis that
the pairing of electrons can also happen provided the frequency
of their oscillations is significantly reduced owing to nonlinear
electron-electron or electron-ion interactions (see, e.g., Refs.~\cite{Zak72,SkoHaa80,DavYakZal05}). However, this issue
requires an additional special study.

We described the formation of pairs of water ions inside a plasmoid. This process may result
in the appearance of superconducting phase in plasma. Note that
the idea that the plasma superconductivity may explain the stability
of a natural plasma structure, called ball lightning (BL)~\cite{BycNikDij10}, was previously discussed
in Refs.~\cite{Dij80,Zel06}. As we have seen above, for
the pairing of charged particles to happen, the typical diameter
of a plasmoid should be very small, $\sim(10^{-7}-10^{-6})\ \text{cm}$.
The plasma structures of the similar size, based on quantum oscillations
of electrons, were studied in Ref.~\cite{DvoDvo06}. It should be noted that previously the magnetic interaction~\cite{Mei84} and the exchange of a virtual Langmuir wave~\cite{Vek12} were considered as possible mechanisms which underlie the plasma superconductivity.

It should be noticed that the existence of the superconducting state of plasma inside BL can result in the presence of strong nondecaying electric currents in this kind of objects. It may explain the fact that sometimes the appearance of a natural BL magnetically
affected heavy metallic objects like a church bell~\cite{Bla73}.

The results of our work can applied for the explanation of the BL creation in natural conditions. It is very difficult to excite a weakly decaying plasma oscillation. However, if we suppose that the pairing of charged particles in a quantum plasmoid, and possibly superconductivity, happens at the initial stages of plasma structure evolution, one can account for the creation of BL in a drop of rain water. Moreover our
results may be used in the interpretation of the experiments with
electric discharges in liquid water~\cite{Gol90,Sha02,Ver08}, where
luminous spherical objects, resembling a natural BL, were generated.

We should also mention that separate tiny plasmoids, each of them being a radial oscillation of charged particles, can form a composite object due to the quantum exchange interaction~\cite{Dvo12JASTP}. Note that the model of a composite BL, which is confirmed by observations~\cite{BycNikDij10}, was also recently discussed in Ref.~\cite{Mes07}.


\section{Conclusion\label{sec:CONCL}}

In conclusion we mention that in the present work we have developed
the theory of quantum spherical plasmoids. Such a plasma structure
is based on radially symmetric oscillations of charged particles in
plasma. The method of linearized quantum hydrodynamics~\cite{Man05,VlaTys11}, which is frequently used to account for quantum
effects in plasmas, seems to be quasi-classical. On the contrary, our method involves
quantum mechanical description of the electrons and ions evolution
in a self-consistent potential which governs the oscillatory motion
of charged particles. Thus our approach is valid for the description
of dense quantum plasmas where quantum effects are important.

We have developed a model of quantum plasmoid which involves the secondly quantized motion of charged particles. Note that the formalism of creation and annihilation operators was also previously used in plasma physics~\cite{Kre05}. In frames of this approach various aspects of quantum plasma dynamics, like phase transitions~\cite{SchBonTsc95}, the behavior under the influence of strong laser fields~\cite{KreBorBon99}, the time evolution in external fields~\cite{BonKwoSemKre99}, and the particle trapping in a strong
electromagnetic field~\cite{FroBonDuf08}, have been studied. The recent applications of the nonequilibrium Green function formalism to the solid state physics can be found in Ref.~\cite{BalBon09}. We also mention an alternative description of quantum plasma dynamics, which is based on Monte Carlo simulations~\cite{FilBonLoz01}. In the present work for the first time the have used the second quantization formalism for the studies of a confined plasma stricture which has a spherical symmetry.

In Sec.~\ref{sec:QUANTEL} we have applied the general formalism
to the quantization of charged particles in a spherically symmetric
plasmoid. In particular we have derived the secondly quantized Hamiltonian~(\ref{eq:Helsq})
and have constructed the ground state of the system. In our work we
have discussed the situation when, along with oscillating charged
particles, there is a neutral component in plasma. In Sec.~\ref{sec:QUANTPH}
we have secondly quantized the field of acoustic waves which are inevitably
excited inside a plasmoid. The quantization of a phonon field is necessary since later we discuss possible applications which involve phonons propagation in dense media, like nuclear matter of NS and liquid water, cf. Sec.~\ref{sec:APPL}.

Then, in Sec.~\ref{sec:INTELACOU}, we
have considered the interaction between charged particles and acoustic
waves. We have derived the secondly quantized Hamiltonian of charged
particles which accounts for the effective interaction owing to the
exchange of virtual acoustic waves, cf. Eq.~(\ref{eq:Htotgen}).
We have also considered the situation when this interaction can be
attractive. In Sec.~\ref{sec:DIAGHAM} we have discussed the attractive
interaction between charged particles in details and have shown that
it can result in the transformation of the ground state of the system.
We have demonstrated that charged particles, which are supposed to
be $1/2$-spin fermions, tend to form singlet bound states which
are more favorable energetically. This phenomenon is analogous to the formation
of Cooper pairs in a metal, that results in the metal superconductivity.

In Sec.~\ref{sec:APPL} we have discussed the possible applications
of the pairing of charged particles in plasma owing to the exchange
of virtual acoustic waves. Firstly, we have considered the case of
a very dense plasma corresponding to the inner crust of NS. It has been shown that protons in NS matter can form pairs. This result is in agreement with the
previous findings that proton superconductivity may well happen in
the NS matter. Secondly, we have studied the possibility
of pairing of singly ionized water ions inside a plasmoid in
the terrestrial conditions. We have shown that this phenomenon can
happen provided a plasma structure is created in a liquid water and
the density of ions is quite high. We have also considered the implication
of our results to the theoretical description of stable natural plasma
objects as well as to the resent experiments where plasmoids were
generated in electric discharged in water.

We have obtained that the pairing is unlikely to occur in the
electron component of plasma. It happens because of the quite high
frequency of electron oscillations. Thus electrons will excite acoustic
waves ineffectively. Moreover we should take into account the higher
electron temperature compared to that of ions. It means that the formation
of bound states of electrons will be washed out by their thermal motion.
This result is consistent with the claim of the authors of Ref.~\cite{VlaYak78}
who found that the exchange of an acoustic wave leads to the cohesion
of ions rather than electrons.

\section*{Acknowledgments}
I am thankful S.~I.~Dvornikov for helpful discussions and to FAPESP
(Brazil) for a grant.

\end{document}